\documentclass[a4paper,superscriptaddress,twocolumn,showkeys,prX]{revtex4-1}
\usepackage{amssymb, amsmath}
\usepackage{graphics}
\usepackage{pst-all}
\usepackage{hyperref}
\usepackage[T1]{fontenc}
\usepackage[utf8]{inputenc}
\usepackage{epsfig}
\newcommand{\ud}{\,\mathrm{d}}

\begin{document}
\title{Phase transitions in the two-dimensional Anisotropic Biquadratic Heisenberg Model}

\author{Antônio R. Moura}
\email{armoura@infis.ufu.br}
\affiliation{Universidade Federal de Uberlândia}

\author{Antônio S. T. Pires}
\affiliation{Universidade Federal de Minas Gerais}
\email{antpires@fisica.ufmg.br}

\author{Afrânio R. Pereira}
\affiliation{Universidade Federal de Viçosa}
\email{apereira@ufv.br}

\date{\today}

\begin{abstract}
In this paper we study the influence of the single-ion anisotropy in the two-dimensional biquadratic Heisenberg model (ABHM) on
the square lattice at zero and finite low temperatures. It is common to represent the bilinear and biquadratic terms by $J_1=J\cos\theta$ and 
$J_2=J\sin\theta$, respectively, and it is well documented the many phases present in the model as function of $\theta$. However we have adopted
a constant value for the bilinear constant ($J_1=1$) and small values of the biquadratic term ($|J_2|<J_1$).
In special, we have analyzed the quantum phase transition due to the single-ion anisotropic constant $D$. For values below a critical anisotropic constant $D_{c}$ 
the energy spectrum is gapless and at low finite temperatures the order parameter
correlation has an algebraic decay (quasi long-range order). Moreover, in $D<D_c$ phase there are a transition temperature
where the quasi long-range order (algebric decay) is lost and the decay becomes exponential, similar to the
Berezinski-Kosterlitz-Thouless (BKT) transition. For $D > D_c$, the excited states are gapped and there is no spin long-range
order (LRO) even at zero temperature. Using Schwinger bosonic representation and Self-Consistent Harmonic
Approximation (SCHA), we have studied the quantum and thermal phase transitions as a function
of the bilinear and biquadratic constants.
\end{abstract}


\maketitle

\section{Introduction}
The subject of low-dimensional magnetic models has received a lot of attention since
Haldane discovered the intriguing physics of the quantum spin chains. In one dimension, spin
models present completely different behavior if the spin value is integer or semi-integer due purely to
quantum effects \cite{PRL61}. While semi-integer spin chains have a gapless energy
spectrum, the corresponding integer spin chain presents the famous Haldane gap. Other
important effects have also been predicted (not only in the condensed matter physics) and
observed firstly in magnetic materials as, for example, the spontaneous broken symmetry\cite{NC19}.
It is well known that the $SO(3)$ spin symmetry is spontaneous broken in the ordered phases
and to restore the lost symmetry, spin-waves (Goldstone modes) emerge in the magnetic background.
Quantum phase transition (QPT) is another pertinent topic in magnetic
systems\cite{sachdev,PRB77}. Differently from thermal phase transitions, the QPT occurs
at zero temperature when a nonthermal parameter is changed.  Many times, the effects of a
QPT can be measured at finite temperatures and so it is supposed that such effects are responsible for
new interesting phenomena. 

Pires \textit{et al.} have studied the critical properties
of the two-dimensional anisotropic XY-model\cite{EPJB44,JPCM20,PA388}. These authors
have shown that a QPT can be governed by the constant associated with the single-ion anisotropy. In the present
paper we have investigated the thermal and quantum phase transitions in the anisotropic
biquadratic Heisenberg model (ABHM) on a two-dimensional square lattice, defined by the following Hamiltonian:
\begin{equation}
\label{main_hamiltonian}
H=\sum_{\langle ij\rangle}\left[J_1({\bf S}_i\cdot{\bf S}_j)+J_2({\bf S}_i\cdot{\bf S}_j)^2\right]+\sum_i D(S_i^z)^2,
\end{equation}
where $J_1$, $J_2$ and $D$ are the bilinear, biquadratic and single-ion anisotropy constants,
respectively. Here we have used a constant value to $J_1$ and expressed $J_2$ and $D$ as function of $J_1$. 
The first sum is taken over the nearest neighbors while the second one is over
all the sites. It is important to note that the above Hamiltonian only makes sense for spins
larger than $1/2$ due to the anisotropic term. The spin-$1/2$ case is degenerated and both up
and down states have the same energy. Obviously a magnetic field could break the degeneracy
separating the energy spectrum. Thus, here we have considered the more relevant case with spin-$1$
where there are two energy bands: $S_i^z=\pm1$ and $S_i^z=0$. The same model on a triangular
lattice has recently been studied by Serbyn \textit{et al.} \cite{PRB84} through a fermionic representation. 
These authors have obtained the phase diagram as well as the specific heat and susceptibility.

The biquadratic term arises from fourth-order perturbations in the exchange interaction
and normally its value is smaller than the bilinear term (although there are cases where
the biquadratic prevails over bilinear interaction). The model studied has a very rich set
of phases as a function of constants $J_1$ and $J_2$ written as  $J_1=J\cos\theta$
and $J_2=J\sin\theta$. For the one dimensional model, the many phases are very well knwon in the literature 
\cite{PRB43,JPCM5,JMMM140,PRB58,PRB74,PRB80}. Indeed, the points $\theta=\pi$ and $\theta=0$
correspond to the pure ferromagnetic (FM) and antiferromagnetic (AFM) models respectively. In the interval $\pi/2<\theta<5\pi/4$
one has a stable ferromagnetic regime with long-range order (LRO); in $-3\pi/4<\theta<-\pi/4$
a dimerized phase arises while $-\pi/4<\theta<\pi/4$ leads to an antiferromagnetic phase with Haldane gap
(for spin-$1$) and for $\pi/4<\theta<\pi/2$ there is a trimerized phase. Some points have an exact solution
as $\theta=\pm\pi/4$, which are resolved by the Bethe ansatz. Already for the two dimensional model, 
Ivanov \textit{et al.} \cite{PRB68,PRB77b} have shown the existence
of a nematic phase for $\theta\gtrsim 5\pi/4$ while for $\theta\lesssim 5\pi/4$ there are
a disordered nematic phase. Rodrígues \textit{et al.} \cite{PRL106} have also determined the many phases of the
biquadratic anisotropic model to Mott insulators (at unit filling). Using an effective field model (similar to that used
by Ivanov \textit{et al.} \cite{PRB68,PRB77b}) and other methods, they have studied the model in the interval $-0.9\pi\lesssim\theta\leq\-0.5\pi$ for many dimensions. For $\theta<-3\pi/4$, there are a $XY$ ferromagnetic state for small positive values of $D$ and a disordered phase for large positive $D$, while for negative values of $D$ one has an Ising-FM state. When $\theta>3\pi/4$ there are a $XY$ nematic phase if $D<0$ and an Ising nematic phase for $D>0$. The case $J1=J2$ in a Mott insulator with $1/3$ filling was also analyzed by Tóth \textit{et al.} \cite{PRL105}. These authors have shown the existence of a three-sublattice long-range order on the square lattice in favor to the two-sublattice (Néel state) at zero temperature while at finite temperatures, the thermal fluctuations stabilize the two-sublattice state.

The single-ion anisotropy (expressed by the constant $D$) has an important role in the model investigated here. It is present in materials as $N_i(C_2H_8N_2)-2NO_2(ClO_4)$  and it is responsible for quantum phase transitions
\cite{PRB48a,PRB49,PRB66,PRB71}. For instance, there are two distinct regions separated by a
critical value $D_c$ of the anisotropic constant, each one with unique properties\cite{PLA360,JPCM5}.
Below the critical point, the system has a gapless energy spectrum  and a quasi long-range
order (LRO) with an algebraic decay for the spin order-parameter correlation at low finite temperature.
Above a critical transition temperature, the order-parameter decays exponentially and there
is no more LRO. This phenomenon is similar to the Berezinsky-Kosterlitz-Thouless (BKT) transition which occurs in the two-dimensional
XY-model, where there are bound and unpaired vortex-antivortex phases separated by a finite
temperature\cite{JPC6}. As $D$ increases, the transition temperature decreases
(it vanishes at zero temperature for $D=D_c$). For values of the anisotropic constant above $D_c$ (the so-called
large-$D$ phase), the system has a different behavior. The ground state
is unique and associated to $S^z=0$ sector while the excited states are gapped, belonging
to $S^z=\pm1$ sector. In the large-$D$ phase, there is no LRO even at zero temperature (the
order is totally lost), while the excitations have spin one and an infinite lifetime at low
energies. The system suffers a quantum phase transition at zero temperature from a gapless
to a gapped energy state. In the present paper we have adopted the antiferromagnetic bilinear model with
$J_1>0$ and small biquadratic constant, $-J_1<J_2<J_1$. Once we are interested in the phase transition for large values
of $D$, we have considered only positive values for the anisotropic constant.

To investigate the behavior of the ABHM, we have applied two different methods at zero temperature:
the first method is used for the case $D<D_c$ and the second for the large-$D$ phase. Below the critical
point, we have used the Schwinger bosonic representation\cite{PRB38,PRB40,PRB45} which indicates a phase transition
close to $D_c$, although it does not provide the exact transition point. Better results are
found analyzing the large-$D$ phase, where we have used the bond operator formalism at zero
temperature developed by L. F. Hai and F. X. Zhi\cite{PLA360}. For finite temperatures, the most appropriate spin-wave method is
the Self-Consistent Harmonic Approximation (SCHA)\cite{JPC7,PA388,PRB48,PRB53,PRB54}.
Using the SCHA, we have determined the transition temperature for many combinations of
the bilinear, biquadratic and anisotropic constants. Extrapolating to zero temperature, the
SCHA also provides the values of the critical points $D_c$, close to that one obtained from
bond operator formalism. In the next section, we present the results for the phase $D<D_c$,
while in section \ref{large-d} we present the results for the large-$D$ phase. The
section \ref{scha} is dedicated to finite temperatures analysis and finally, the conclusion is
given in section \ref{conclusion}.

\section{Schwinger bosonic representation}
\label{smaller-d}

\begin{figure}[h]
\centering
\epsfig{file=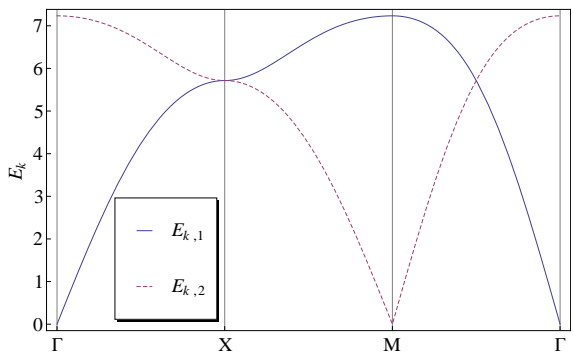,height=50mm}
\caption{The energy spectrum $E_{{\bf k},1}$ and $E_{{\bf k},2}$. Here $\Gamma=(0,0)$, $X=(\pi,0)$ and $M=(\pi,\pi)$. }
\label{fig:energies}
\end{figure}

In the phase with $D<D_c$, the ground state is ordered at zero temperature and without LRO at
finite temperatures as dictated by the Mermin-Wagner theorem\cite{PRL17}. The lowest energy
excitations are the gapless Goldstone bosons which emerge with any amount of energy. We have
used the $SU(2)$ Schwinger bosonic formalism to describe this phase. The spin in each site
$i$ is represented by two bosonic operators $a_i$ e $b_i$. Forgetting the anisotropic term for
a while, the action for the model is given by:
\begin{equation}
\mathcal{Z}=\int\mathcal{D}[{\bf S}_i]e^{-i\int H \ud t},
\end{equation}
where:
\begin{eqnarray}
\label{decoupled_hamiltonian}
H&=&\sum_{\langle i,j\rangle}\left[2J_2\langle{\bf S}_i\cdot{\bf S}_j\rangle({\bf S}_i\cdot{\bf S}_j)-J_2\langle{\bf S}_i\cdot{\bf S}_j\rangle^2
+\right.\nonumber\\
&&\left.+J_1({\bf S}_i\cdot{\bf S}_j)\right].
\end{eqnarray}
In the above equations we have applied the Hubbard-Stratonovich transform\cite{SPD2,PRL3} to decouple
the biquadratic term in favor to the mean-field parameter $\langle {\bf S}_i\cdot{\bf S}_j\rangle$.
In the limit $J_2=0$, we recover the traditional Heisenberg model. Because the sucessive mean-field approximation, 
this method is not the most suitable one and some results are more qualitative than quantitative. Accurate 
results are given in the next section. The mean-field parameter will
be determined by the minimum of the Helmholtz free energy. The spin operators are written as
$S_i^+=a_i^{\dagger}b_i$, $S_i^-=b_i^{\dagger}a_i$ and $S_i^z=(a_i^{\dagger}a_i-b_i^{\dagger}b_i)/2$,
where $a_i$ and $b_i$ are the two Schwinger bosons on site $i$. The bosonic operators keep the
spin commutation relation and to ensure $S^2_i=S(S+1)$ we have to impose the local constraint
$\sum_i(a_i^\dagger a_i+b_i^\dagger b_i)=2S$, which fixes the total number of bosons on each site.
Therefore, in the bosonic formalism, the bilinear term is written as
\begin{equation}
{\bf S}_i\cdot{\bf S}_j=-\frac{1}{2}\mathcal{A}_{ij}^{\dagger}\mathcal{A}_{ij}+S^2,
\end{equation}
with the bond operator $\mathcal{A}_{ij}=a_ia_j+b_ib_j$ (as usually we make a rotation by $\pi$ around
the $y$-axis on sublattice B), while the biquadratic interaction is given by:
\begin{equation}
\sum_i(S_i^z)^2=\frac{1}{4}\sum_i(a_i^{\dagger}a_i-b_i^{\dagger}b_i)^2=-\sum_i a_i^{\dagger}b_i^{\dagger}a_ib_i,
\end{equation}
where we used the constraint $\sum_i(a_i^\dagger a_i+b_i^\dagger b_i)=2S$ to simplify the
expression and a constant term was discarded by a redefinition of the ground state energy.
Both interactions, the bilinear and the biquadratic ones, are fourth order terms in the action
and they are decoupled by using the Hubbard-Stratonovich transform again:
\begin{equation}
\mathcal{A}_{ij}^{\dagger}\mathcal{A}_{ij}\rightarrow -A(\mathcal{A}_{ij}^{\dagger}+\mathcal{A}_{ij})-A^2
\end{equation}
and
\begin{equation}
a_i^{\dagger}b_i^{\dagger}a_i b_i\rightarrow B (a_i b_i+a_i^{\dagger}b_i^{\dagger})-B^2,
\end{equation}
where we have introduced the real mean-fields
$A=\langle\mathcal{A}_{ij}^{\dagger}\rangle=\langle\mathcal{A}_{ij}\rangle$ and
$B=\langle a_i^{\dagger}b_i^{\dagger}\rangle =\langle a_ib_i\rangle$. The values of $A$ and
$B$ are also determined by minimizing the Helmholtz free energy. Thus, the second-order
Hamiltonian to the ABHM is:
\begin{eqnarray}
H&=&H_0+\sum_i\left[\frac{1}{2}\lambda(a_i^{\dagger}a_i+b_i^{\dagger}b_i)-BD(a_ib_i)+h.c.\right]+\nonumber\\
&&+\tilde{A}\sum_{\langle i,j\rangle}(a_i a_j+b_i b_j+h.c.)
\end{eqnarray}
with the constant term
\begin{eqnarray}
H_0&=&\left(\frac{J_1}{2}A^2+J_2S^2A^2-\frac{3J_2}{4}A^4\right)\frac{Nz}{2}+NDB^2-\nonumber\\
&&-2N\lambda \left(S+\frac{1}{2}\right)
\end{eqnarray}
and $\tilde{A}=-\left(\frac{J_1}{2}A+J_2S^2A-\frac{J_2}{2}A^3\right)$. Here, a constraint has been
added by a local Lagrange multiplier $\lambda_i$ on each site and after that we have adopted
a mean value $\lambda=\langle\lambda_i\rangle$. According to Auerbach \cite{PRB38}, the use of a
mean value for $\lambda$ causes an incorrect prediction of ${\bf S}^2$ and some related quantities.
The predict value is smaller than the correct one by a factor of $3/2$. The correction can
be done by a perturbation expansion around the mean value but we have used the ordinary method
of adding a $3/2$ factor when it is necessary. Takahashi has shown that this factor does not appear in
a theory based on the Holstein-Primakoff representation\cite{PRB36,PRB40b}. He has naturally
obtained the same corrected equations as in the Schwinger formalism; however, this work is substantially
simpler in the Schwinger representation and to our purposes, this is an acceptable method. After a
Fourier transform, the Hamiltonian in momentum space is given by:
\begin{equation}
H=H_0+\frac{1}{2}\sum_{\bf k} \beta_{\bf k}^{\dagger}H_{\bf k}\beta_{\bf k},
\end{equation}
where the vector $\beta_{\bf k}^{\dagger}=(a_{\bf k}^{\dagger}\ \ b_{\bf k}^{\dagger}\ \ a_{-{\bf k}}\ \ b_{-{\bf k}})$
and the matrix is:
\begin{equation}
\tilde{H}=\left(\begin{array}{cccc}
\lambda & 0 & 4\tilde{A}\gamma_{\bf k} & -BD\\
0 & \lambda & -BD & 4\tilde{A}\gamma_{\bf k}\\
4\tilde{A}\gamma_{\bf k} & -BD & \lambda & 0\\
-BD & 4\tilde{A}\gamma_{\bf k} & 0 & \lambda\\
\end{array}\right),
\end{equation}
in which $\gamma_{\bf k}=\sum_{\delta}e^{i{\bf k} \cdot{\bf r}_{\delta}}=\frac{1}{2}(\cos k_x+\cos k_y)$
is the structure factor and $\delta$ designates the four neighboring sites (we have assumed an unitary
lattice parameter, $a=1$). The diagonalization is done by using the Bogoliubov method in order
to keep the bosonic nature of the operators. The energy eigenvalues are given by
$E_{{\bf k},1}=\sqrt{\lambda^2-(4\tilde{A}\gamma_{\bf k}+BD)^2}$ and $E_{{\bf k},2}=\sqrt{\lambda^2-(4\tilde{A}\gamma_{\bf k}-BD)^2}$
while the Hamiltonian assumes the quantum harmonic oscillator structure:
\begin{equation}
H=H_0+\sum_{{\bf k},m}\left(c_{{\bf k}m}^{\dagger}c_{{\bf k}m}+\frac{1}{2}\right)E_m,
\end{equation}
in which $c_{{\bf k}m}$ ($m=1,2,3$) are the new bosonic operators. The Helmholtz free energy
($F=-\beta^{-1}\ln\textrm{Tr} e^{-\beta H}$) is therefore:
\begin{equation}
F=H_0+\frac{1}{\beta}\sum_{{\bf k},m}\left\{\ln\left[\textrm{sinh}\left(\frac{\beta E_m}{2}\right)\right]\right\},
\end{equation}
where the sum is taken over the first Brioullin zone. By optimization of the Helmholtz free energy,
$\partial F/\partial A=\partial F/\partial B=\partial F/\partial\lambda=0$, we have obtained
three integral self-consistent equations. In the continuous limit and at zero temperature, they become:
\begin{subequations}
\begin{eqnarray}
&A=-\frac{1}{2}\int\frac{\ud^2{\bf k}}{4\pi^2}\left[\frac{4\tilde{A}\gamma_{\bf k}+BD}{E_{{\bf k},1}}+
\frac{4\tilde{A}\gamma_{\bf k}-BD}{E_{{\bf k},2}}\right]\gamma_{\bf k},\\
&B=\frac{1}{4}\int \frac{\ud^2{\bf k}}{4\pi^2}\left[\frac{4\tilde{A}\gamma_{\bf k}+BD}{E_{{\bf k},1}}-
\frac{4\tilde{A}\gamma_{\bf k}-BD}{E_{{\bf k},2}}\right],\\
&S+\frac{1}{2}=\frac{1}{4}\int\frac{\ud^2{\bf k}}{4\pi^2}\left[\frac{\lambda}{E_{{\bf k},1}}+\frac{\lambda}{E_{{\bf k},2}}\right].
\end{eqnarray}
\end{subequations}
Obviously, an exact solution for $A$, $B$ and $\lambda$ mean-field parameters is a hard task
to be obtained and numerical methods must be used. However, we can have a general idea of the
system behavior from the existence or not of solutions. According to Takahashi and Arovas,
at finite temperatures, the self-consistent equations have a solution and it is associated with
a phase with vanishing spin order. The ground state is disordered and the exponential decay of the spin-spin correlation
at large distances can be demonstrated explicitly.  At zero temperature, there is no
solution to the integral equations because the bosons condensate in a state with zero energy;
the equations are divergent. The absence of solution is a feature of an ordered ground
state, i.e. a state with broken spin symmetry. Similar to the Bose-Einstein condensation, the
divergence is countered by separating the term with zero energy from the integral equations. In our
case, $E_{{\bf k},1}$ vanishes at point ${\bf k}^\ast=(\pm\pi,\pm\pi)$, while $E_{{\bf k},2}$ reaches the minimum at
${\bf k}^\ast=(0,0)$; so $\lambda=|4\tilde{A}\gamma_{{\bf k}^\ast}+BD|$ in the condensate
state. Close to the minimum energy point, we have a massless relativistic dispersion relation
$E_{\bf k}=|{\bf k-k}^\ast| c$, where $c$ is the spin-wave velocities. The excited states are the gapless
Goldstone modes that emerge from any amount of energy in order to try to restore the broken spin symmetry. In the Figure (\ref{fig:energies})
we shown the energy spectrum of $E_{{\bf k},1}$ and $E_{{\bf k},2}$ for $J_1=1$, $J_2=0.25 J_1$ and $D=4J_1$.

\begin{figure}[h]
\centering
\epsfig{file=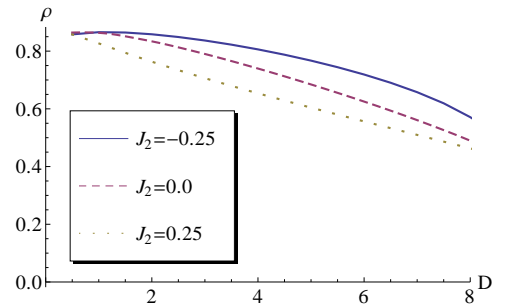,height=50mm}
\caption{The condensate density as a function of the anisotropic constant $D$ (in units of $J_{1}$). For $D>1$, the
density decays almost linearly and it is expected that $\rho=0$ at a critical point $D_c$.}
\label{fig:rho_D}
\end{figure}

\begin{figure}
\centering
\epsfig{file=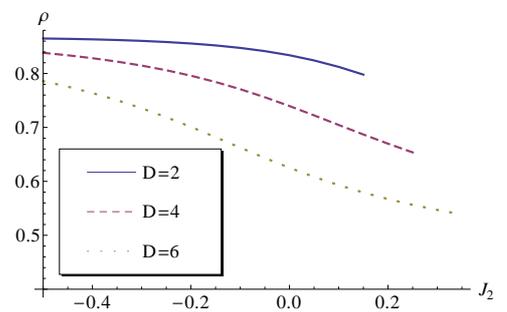,height=50mm}
\caption{The condensate density as a function of the biquadratic constant $J_2$ (in units of $J_{1}$).
There ia only a small influence for the $J_2$ considered values.}
\label{fig:rho_J2}
\end{figure}

After separating the divergent term of the integral equations, we introduce a new parameter $\rho$
that measures the condensate density as following:
\begin{equation}
\rho=\left(S+\frac{1}{2}\right)-\frac{1}{4}\int\frac{\ud^2{\bf k}}{4\pi^2}\left(\frac{\lambda}{E_{{\bf k},1}}+\frac{\lambda}{E_{{\bf k},2}}\right).
\end{equation}
Therefore the equations for $A$ and $B$ are given by:	
\begin{eqnarray}
&A=-2\rho-\frac{1}{2}\int\frac{\ud^2{\bf k}}{4\pi^2}\left(\frac{4\tilde{A}\gamma_{\bf k}+BD}{E_{{\bf k},1}}+\frac{4\tilde{A}\gamma_{\bf k}-BD}{E_{{\bf k},2}}\right)\gamma_{\bf k}
\end{eqnarray}
and
\begin{eqnarray}
&B=\rho+\frac{1}{4}\int\frac{\ud^2{\bf k}}{4\pi^2}\left(\frac{4\tilde{A}\gamma_{\bf k}+BD}{E_{{\bf k},1}}-\frac{4\tilde{A}\gamma_{\bf k}-BD}{E_{{\bf k},2}}\right).
\end{eqnarray}
Now the self-consistent equations can be numerically solved. The results for the condensate
density $\rho$ are shown in Figures (\ref{fig:rho_J2}) and (\ref{fig:rho_D}). The magnetization is proportional to
density $\rho$ so the descrease in the boson condensate implies a lower magnetization. 

In both graphics the anisotropic and biquadratic constants are given in units of $J_1$. The bosons are highly
condensed to small values of the anisotropic constant $D$. At $D=0$, the level is around 82$\%$
for different biquadratic constants but close to $D=8$ it is approximately 60$\%$, while it does not
suffer a notable influence for small values of the biquadratic interaction constant $J_2$. For the Heisenberg
model ($J_2=D=0$) the condensate density is around 81$\%$ so our results are according. 
For $D>2$ the condensate density decays almost linearly for all $J_2$
values analyzed. When $\rho=0$, the system exists in the condensate phase and enters in regime
with vanishing magnetization. The bosons do not condensate anymore in a null energy state and the spectrum becomes gapped
at zero temperature. 

Figure (\ref{fig:rho_D}) suggests us that exist a point where occurs a
quantum phase transition from the ordered state to the unbroken spin symmetry phase at zero temperature.
If we extrapolate the results for $\rho\to0$ and by assuming the linear behavior, we could discover the critical point but
due mean-field approximation adopted, the results are bigger than expected. For large $D$ anisotropic constant the 
condensate level decreases sufficiently to invalidate the initial assumption of a high boson condensation. 
Therefore the theory is not appropriate when the 
condensate density is much lower than $1$ and the results at $D\approx D_c$ are not accurate. 
A better method to obtain the critical point is shown in the next section where we begin in the gapped
phase, with $D>D_c$, and lowered the anisotropic constant to the gapless region. In the Figure (\ref{fig:rho_J2})
we have the density as function of $J_2$ for three values of $D$. As expected, the condensate is small for bigger values 
of the anisotropic constant. 
\begin{figure}[h]
\centering
\epsfig{file=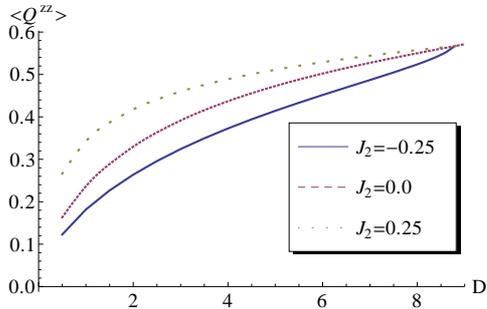,height=50mm}
\caption{The mean value of the quadrupole moment $Q^{zz}$ as function of $D$ below the critical point $D_c$.}
\label{fig:qzz}
\end{figure}

Already in Figure (\ref{fig:qzz}) we shown the quadrupole moment $\langle Q^{zz}\rangle$ as function of 
the anisotropic constant for differents values of $J_2$. The behavior here is distinct from previous (magnetization $S_z$) 
and it increases for large $D$ values. The non vanishing values for the quadrupole moment, defined by $Q^{zz}=2/3-S^z S^z$, indicate
the presence of a nematic phase characterized by a broken spin-rotational symmetry but with a preserverd time-reversal symmetry. 
Therefore, although the magnetization (spin order) goes to zero in the large-D phase, the nematic order increases with the
anisotropic constant.

\section{Large-D phase}
\label{large-d}
\begin{figure}[h]
\centering
\epsfig{file=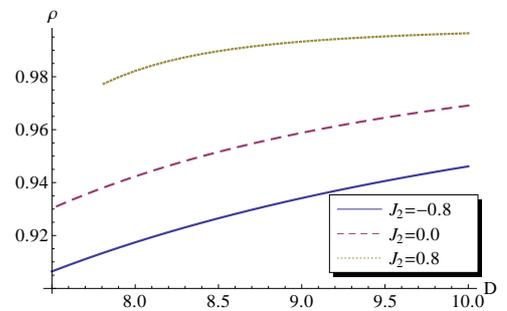,height=50mm}
\caption{The condensate density as a function of the anisotropic constant $D$
(in units of $J_1$) in the large-D phase.}
\label{fig:rho0_bond}
\end{figure}

For anisotropic constant bigger than a critical value $D_c$, the physical properties are
distinct from those studied in previous section. In the so-called large-$D$ phase, the
energy spectrum is gapped and the spin symmetry is restored for a disordered state with $m=0$. A phase
transition occurs at point $D_c$ even at zero temperature, which characterizes a quantum
phase transition. In this section, we analyze the large-$D$ phase by using a $SU(3)$ Schwinger
bosonic representation, the so-called bond operator. The formalism is the same developed in the
reference \cite{PLA360} although we have adopted $J_1=1$ and $|J_2|<J_1$ 
while in the reference the authors have considered $J_1=J\cos\theta$ and $J_2=J\sin\theta$.
The critical anisotropic constant $D_c$ is found as a function of the
biquadratic constant $J_2$ (in units of $J_1$). We begin representing the eigenstates of $S^z_i$ as a
function of three boson operators: $|m_i=-1\rangle=a_{i,-1}^{\dagger}|0\rangle$, $|m_i=0\rangle=a_{i,0}^{\dagger}|0\rangle$
and $|m_i=+1\rangle=a_{i,+1}^{\dagger}|0\rangle$, where $|0\rangle$ is the vacuum state.
The boson operator $a^\dagger_{im}$ creates a particle with $m_z=m$ on site $i$. The commutation
relations $[S_i^+,S_j^-]=2S_i^z\delta_{ij}$ and $[S_i^z,S_j^{\pm}]=\pm S_i^{\pm}\delta_{ij}$ are
valid and to keep $S^2_i=S(S+1)$ we have to impose again a constraint $\sum_\mu a^\dagger_{i,\mu} a_{i,\mu}=S$
on each site. A condensation occurs in the $|m=0\rangle$ state, once this is the smaller
band energy (the $|m=\pm 1\rangle$ states are degenerate). Therefore the number of particles is
$N_0=\langle a^\dagger_{i,0} a_{i,0}\rangle\gg 1$ and we can consider the approximation
$[N_0,a^\dagger_{i,0}]=0$ and $[N_0,a_{i,0}]=0$, which allows to treat the $a_{i,0}$ operators as
real numbers. Thus we have adopted the mean values
$\langle a^\dagger_{i,0}\rangle=\langle a_{i,0}\rangle=a_0$  in the next equations. In the $SU(3)$
bosonic representation, the bilinear and biquadratic spin interactions are written as:
\begin{widetext}
\begin{eqnarray}
({\bf S}_i\cdot{\bf S}_j)&=&a_0^2\left(a_{i,-1}a_{j,1}+a_{i,-1}a_{j,-1}^{\dagger}+a_{i,1}^{\dagger}a_{j,1}+a_{i,1}^{\dagger}a_{j,-1}^{\dagger}+a_{i,1}a_{j,-1}+a_{i,1}a_{j,-1}^{\dagger}+
a_{i,-1}^{\dagger}a_{j,-1}+a_{i,-1}^{\dagger}a_{j,1}^{\dagger}\right)+\nonumber\\
&&+\left(a_{i,1}^{\dagger}a_{i,1}a_{j,1}^{\dagger}a_{j,1}-a_{i,1}^{\dagger}a_{i,1}a_{j,-1}^{\dagger}a_{j,-1}-a_{i,-1}^{\dagger}a_{i,-1}a_{j,1}^{\dagger}a_{j,1}+a_{i,-1}^{\dagger}a_{i,-1}a_{j,-1}^{\dagger}a_{j,-1}\right),
\end{eqnarray}
\end{widetext}
and
\begin{widetext}
\begin{eqnarray}
({\bf S}_i\cdot{\bf S}_j)^2&=&\left(a_{i,-1}^{\dagger}a_{i,-1}a_{j,1}^{\dagger}a_{j,1}+a_{i,1}^{\dagger}a_{i,1}a_{j,-1}^{\dagger}a_{j,-1}+
a_{i,-1}^{\dagger}a_{j,1}^{\dagger}a_{i,1}a_{j,-1}+a_{i,1}^{\dagger}a_{j,-1}^{\dagger}a_{i,-1}a_{j,1}\right)+(1+a_0^4)-\nonumber\\
&&-a_0^2\left(a_{i,1}^{\dagger}a_{j,-1}^{\dagger}+a_{i,1}^{\dagger}a_{j,1}^{\dagger}+a_{i,1}a_{j,-1}+a_{i,-1}a_{j,1}\right),
\end{eqnarray}
\end{widetext}
where $\mathcal{A}_{ij}=(a_{i,-1}a_{j,1}+a_{i,1}a_{j,-1}-a_0^2)$, while the anisotropic term is:
\begin{equation}
(S_i^z)^2=\left(a_{i,1}^{\dagger}a_{i,1}-a_{i,-1}^{\dagger}a_{i,-1}\right)^2=1-a_0^2.
\end{equation}
The Hamiltonian is composed by fourth-order terms and we have applied again the
Hubbard-Stratonovich transform. After decoupling, we obtain the following
second-order Hamiltonian:
\begin{widetext}
\begin{eqnarray}
\label{eq:hamiltonian_larged}
H&=&(1-\rho_0)ND+(1+\rho_0^2)J_2\frac{Nz}{2}+N\lambda(\rho_0-S)+J_1(1-\rho_0)^2\frac{Nz}{2}-
-4\Lambda^2\left(J_2-J_1\right)\frac{Nz}{2}+\lambda\sum_i\left(a_{i,1}^{\dagger}a_{i,1}+a_{i,-1}^{\dagger}a_{i,-1}\right)+\nonumber\\
&&+\sum_{\langle i,j\rangle}\left[\rho_0\left(a_{i,1}^{\dagger}a_{j,1}+a_{i,-1}^{\dagger}a_{j,-1}+a_{i,1}a_{j,1}^{\dagger}+a_{i,-1}a_{j,-1}^{\dagger}\right)\right.-\rho_0(J_2-J_1)\left(a_{i,1}^{\dagger}a_{j,-1}^{\dagger}+a_{i,-1}^{\dagger}a_{j,1}^{\dagger}+a_{i,1}a_{j,-1}+a_{i,-1}a_{j,1}\right)+\nonumber\\
&&\left.+2(J_2-J_1)\Lambda\left(a_{i,1}^{\dagger}a_{j,-1}+a_{i,-1}^{\dagger}a_{j,1}+a_{i,1}a_{j,-1}+a_{i,-1}a_{j,1}\right)\right],
\end{eqnarray}
\end{widetext}
in which we have introduced  the real mean-field $\Lambda=\langle a_{i,1}a_{j,-1}\rangle=\langle a_{i,1}^{\dagger}a_{j,-1}^{\dagger}\rangle$
and defined the level condensate $\rho_0=a_0^2$.
Performing a Fourier transform, we obtain in the momentum space:
\begin{equation}
H=H_0+\frac{1}{2}\sum_{\bf k} \beta_{\bf k}^{\dagger}\tilde{H}\beta_{\bf k},
\end{equation}
where $H_0$ are the constant term of equation (\ref{eq:hamiltonian_larged}) and the $\tilde{H}$ matrix is given by:
\begin{equation}
\tilde{H}=(\lambda+4\rho_0 J_1\gamma_{\bf k})I_{4\times4} + (4J^\prime\gamma_{\bf k}) M_{4\times4}
\end{equation}
where $I$ is the identity matrix, $M$ is the anti-diagonal matrix,
$J^\prime=(J_1-J_2)(\rho_0-2\Lambda)$ and $\gamma_{\bf k}$ is the
structure factor, identical to that given in the previous section. The eigenvalues of the
above equation are $E_{\bf k}=\sqrt{(\lambda+4J_1\rho_0\gamma_{\bf k})^2-(4J^\prime \gamma_{\bf k})^2}$.
The energy has a minimum at point ${\bf k}^\ast=(\pm\pi,\pm\pi)$ and close to this, we have a
dispersion relation $E_{\bf k}^2=m^2 c^4+|{\bf k}-{\bf k}^\ast|^2 c^2$, where $m$ is the mass of the
excitations and $c$ is the spin-wave velocities. Unlike the region in which $D<D_c$, in the
large-$D$ phase, the energy gap $\Delta=mc^2$ is not null and so, there is no long-range order even
at zero temperature and the magnetization is zero in this phase. The $\rho_0$, $\lambda$ and
$\Lambda$ parameters are determined by the self-consistent equations obtained from the minimum
of the Helmholtz free energy. At zero temperature they are given by:
\begin{subequations}
\begin{eqnarray}
\rho_0&=&2-\int \frac{\ud^2{\bf k}}{4\pi^2}\frac{\lambda+4J_1\rho_0\gamma_{\bf k}}{E_{\bf k}},\\
\Lambda&=&-2\int\frac{\ud^2{\bf k}}{4\pi^2}\frac{(J_1-J_2)\rho_0-2\Lambda(J_1-J_2)}{E_{\bf k}}\gamma_{\bf k}^2,\\
\lambda&=&-4\int\frac{\ud^2{\bf k}}{4\pi^2}\frac{(\lambda+4J_1\rho_0\gamma_{\bf k})J_1-(4J^\prime \gamma_{\bf k})(J_1-J_2)}{E_{\bf k}}\gamma_{\bf k}+\nonumber\\
&&+D+4J_1-4\rho_0(J_1+J_2),
\end{eqnarray}
\end{subequations}
where we applied the continuous limit and the integrals are evaluated over the first
Brillouin zone $-\pi<k_x,k_y<\pi$. At finite temperatures, the integrand is multiplied by
$\coth\left(\frac{\beta E_{\bf k}}{2}\right)=2\left(n_{\bf{k}}+\frac{1}{2}\right)$, where
$n_{\bf k}=(e^{-\beta E_{\bf k}}-1)^{-1}$ is the Bose-Einstein distribution. The analysis at finite
temperature is done in the next section, where we have used the SCHA, so for while we
have considered only the $T=0$ case. Once the lowest energy is finite, there is
no divergence in the above equations and they can be directly solved. The $\rho_0$
term measures the boson condensation in the $|m=0\rangle$ state and it is expected a high level
($\rho_0\approx 1$) in the large-$D$ phase followed by a decreasing near the critical
value $D_c$. On the other hand, the mean-field parameter $\Lambda$ is close to zero in the large-$D$
phase and it increases for $D\approx D_c$. Figure (\ref{fig:rho0_bond}) shows the
$\rho_0$ parameter for some biquadratic constants as a function of $D$ in the large-$D$ phase.
At $D=D_c$, $\rho_0$ presents a discontinuity and there is no more condensation in $|m=0\rangle$
state. For $D<D_c$, the system goes to a gapless region with an ordered ground state as
described in the previous section.

We can use the self-consistent equation to estimate analytically
the transition point $D_c$ as a function of the biquadratic constant. Following
Ref. \cite{PLA360}, we adopt $\Lambda\approx 0$ and thus:
\begin{subequations}
\begin{eqnarray}
\rho_0&=&2-\int\frac{\ud {\bf k}^2}{4\pi^2}\frac{1}{\sqrt{1-\Gamma_{\bf k}^2}},\\
D&=&\int\frac{\ud {\bf k}^2}{4\pi^2}\frac{4J_1\gamma_{\bf k}-\frac{1}{g}+4(J_1-J_2)+4J_2\gamma_{\bf k}\Gamma_{\bf k}}{\sqrt{1-\Gamma_{\bf k}^2}}-\nonumber\\
&&-\frac{2}{g}-8J_2,
\end{eqnarray}
\end{subequations}
in which the dimensionless ratios were defined as $g=\frac{\rho_0}{\lambda}$ and
$\Gamma_{\bf k}=\frac{4(J_1-J_2)g\gamma_{\bf k}}{1+4gJ_1\gamma_{\bf k}}$. At the transition point,
$E_{\bf k}=0$ and, therefore, $\Gamma_{\bf k}=\pm1$. The critical ratio $g$ is $g_c=\frac{1}{4(2J_1-J_2)}$
for $J_2<0.5J_1$ and $g_c=\frac{1}{4J_2}$ for $J_2>0.5J_1$. Thus we have found the following
equation for $D_c$:
\begin{eqnarray}
\label{eq:dc}
D_c&=&\int\frac{\ud {\bf k}^2}{4\pi^2}\frac{4J_1\gamma_{\bf k}-4J_2-g_c^{-1}-4(J_1-J_2)\gamma_{\bf k}\Gamma_{\bf k}}{\sqrt{1-\Gamma_{\bf k}^2}}+\nonumber\\
&&+\frac{2}{g_c}+8J_2.
\end{eqnarray}
Although the above equation provides a simple method to determine $D_c$, it is only a first
approximation. For $|D-D_c|\approx 0$, the condensate level $\rho_0$ becomes smaller than $1$ and hence,
$\Lambda\approx 0$ is not a good consideration. In the large-$D$ phase, the condensation is almost
total and decreases close to $D_c$ (the graphic is not valid for $D<D_c$). The better results are
obtained analyzing the point where the gap vanishes for different values of $J_2$. We begin with
a large anisotropic constant (approximately $10J_1$), decreasing its value until the gapless
phase. The point where the gap vanishes is taken as $D_c$. 
\begin{figure}[h]
\centering
\epsfig{file=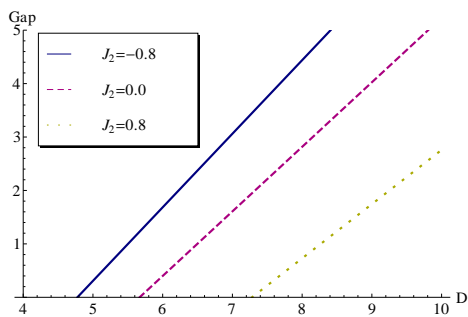,height=50mm}
\caption{The gap energy as a function of $D$.}
\label{fig:gap_d}
\end{figure}

\begin{figure}
\epsfig{file=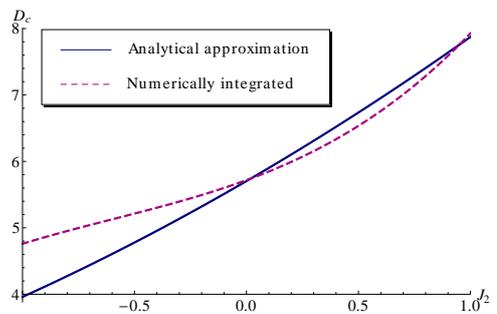,height=50mm}
\caption{The critical points $D_c$ as a function of $J_2$.}
\label{fig:d_j2}
\end{figure}

In Fig. (\ref{fig:gap_d}), we show the gap energy as a function of the anisotropic constant 
$D$ and in Fig. (\ref{fig:d_j2}) the critical points $D_c$ as a function
of the biquadratic constant $J_2$ (all constants are given in units of $J_1$), obtained by numerical integration of
the self-consistent equations and through the approximate analytical method described by equation (\ref{eq:dc}). 
We have evaluated the quantum phase transition in the interval
$-1\leq J_2\leq 1$ (in units of $J_1$) and both results are similar, presenting an almost linear behavior
for $D_c$. The difference between the numeric and approximate analytical results is smaller for
$J_2\approx 0$ where the consideration $\Lambda\approx 0$ is valid and drastically increases
for negative values of the biquadratic constant. As argued earlier, Eq. (\ref{eq:dc}) is only
an approximation and the numeric results are closer to those obtained by SCHA (see next section).

\begin{figure}
\epsfig{file=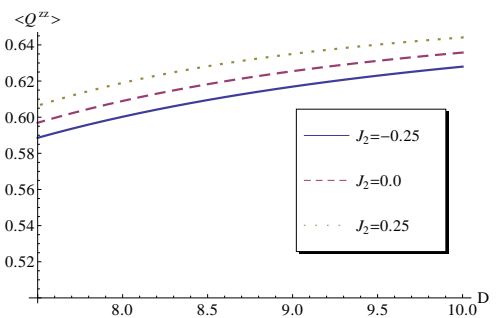,height=50mm}
\caption{The mean value of the quadrupole moment $Q^{zz}$ as function of $D$ in the large-D phase.}
\label{fig:qzzlargeD}
\end{figure}

In the Figure (\ref{fig:qzzlargeD}) we plot the quadrupole moment $\langle Q^{zz} \rangle$ as function of the anisotropic constant. 
Again while the magnetization is nearly zero in the large-D phas,  the quadrupole moment is finite and the results are close those
shown in previous section (small-D phase). Although the system does not present a spin order, the $O(3)$ symmetry is broken by the
nematic order.

\section{Self-consistent Harmonic Approximation}
\label{scha}
Now we analyze the Anisotropic Biquadratic Heisenberg Model (ABHM) by using the Self-consistent Harmonic
Approximation\cite{JPC7,PA388,PRB48,PRB53,PRB54}.
This is the most appropriate method to treat the system at finite temperatures. The
the bosonic formalisms used in the last sections provide reasonable results at zero temperature but,
as argued by Yoshida\cite{JPSJ58}, for $T>0$ they present divergences with another well known results.
As we have demonstrated, the model studied has a QPT associated with the single-ion anisotropy
$D$. There are a critical value $D_c$ in which one separates an spin ordered phase ($D<D_c$) from a
disordered ($D>D_c$) phase.  At finite temperatures the system is always disordered as dictated
by the Mermin-Wagner theorem; however, there are a thermal phase transition from a disordered
state with algebraic decay to a disordered state with exponentially decay for the spin-spin correlation
order-parameter. This is a transition phase, similar to the BKT transition present on the planar
magnets. Using the SCHA, we have determined both quantum and thermal transition to the ABHM.

\begin{figure}[h]
\centering
\epsfig{file=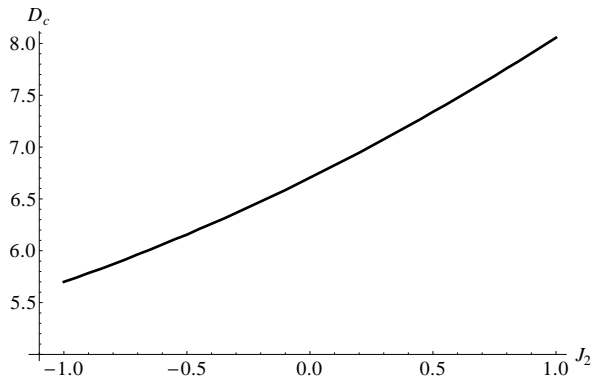,height=50mm}
\caption{The critical values $D_c$ as a function of the biquadratic constant $J_2$ in units of $J_1$.}
\label{fig:dc_scha}
\end{figure}

Starting with Hamiltonian (\ref{main_hamiltonian}), we decouple the biquadratic term by using the
Hubbard-Stratonovich transform $({\bf S}_i\cdot{\bf S}_j)^2\to 2\xi ({\bf S}_i\cdot{\bf S}_j)$,
where $\xi=\langle{\bf S}_i\cdot{\bf S}_j\rangle$ measures the correlation between the nearest
neighbor sites. Therefore, the uncoupled Hamiltonian is given by:
\begin{equation}
\label{eq:original_hamiltonian}
H=\frac{1}{2}\sum_{\langle i,j\rangle} (J_1+2J_2\xi){\bf S}_i\cdot{\bf S}_j+D\sum_i (S_i^z)^2,
\end{equation}
in which $i$ runs over the square lattice and $j$ indicates the four nearest neighboring spins
(the half-integer factor is included to avoid double counting). Following the standard procedures,
we written the spin operator using the Villain's representation:
\begin{subequations}
\begin{eqnarray}
S_i^+&=&e^{i\phi_i}\sqrt{S(S+1)-S_i^z(S_i^z+1)}\\
S_i^-&=&\sqrt{S(S+1)-S_i^z(S_i^z+1)}e^{-i\phi_i},
\end{eqnarray}
\end{subequations}
where $\phi_i$ is the angle of the parametrization:
\begin{footnotesize}
\begin{equation}
{\bf S}_i=(-1)^i\left(\tilde{S}\sqrt{1-\left(\frac{S_i^z}{\tilde{S}}\right)^2}\cos\phi_i,\tilde{S}\sqrt{1-\left(\frac{S_i^z}{\tilde{S}}\right)^2}\sin\phi_i,S_i^z\right)
\end{equation}
\end{footnotesize}
with $\tilde{S}=\sqrt{S(S+1)}$. In order to avoid divergences, we choose the angle operator $\phi_i$
relative to the direction of the instantaneous total spin ($\langle\phi_i\rangle$ is not well
defined to angles measured relative to a fixed axis). Thus, we make the replacement:
\begin{eqnarray}
{\bf S}_i\cdot{\bf S}_j&\to&-\frac{\tilde{S}^2}{2}\sqrt{1-\left(\frac{S_i^z}{\tilde{S}}\right)^2}\sqrt{1-\left(\frac{S_j^z}{\tilde{S}}\right)^2}\cos(\phi_i-\phi_j)+\nonumber\\
&&+S_i^zS_j^z.
\end{eqnarray}
At low temperatures (in the ordered phase), the spin field assumes a configuration with a small
angular difference between neighboring sites and, hence, we can consider $|\phi_i-\phi_j|\ll 1$.
Therefore, expanding the above equation into powers of $(S_i^z/\tilde{S})^2$ and
$(\phi_i-\phi_j)^2$, we have the quadratic Hamiltonian:
\begin{eqnarray}
\label{harmonic_hamiltonian}
H&=&\frac{g}{2}\sum_{\langle i,j\rangle}\left[\xi\tilde{S}^2(\phi_i^2-\phi_i\phi_j)+(S_i^z)^2+S_i^z S_j^z\right] \nonumber\\
&&+ D\sum_i (S_i^z)^2,
\end{eqnarray}
where $g=(J_1+2J_2\xi)$. The $\xi$ parameter inserted before the $\phi$ operators takes into
account non harmonic terms neglected when the original Hamiltonian is written in the quadratic
form\cite{PRB48,ZETF65,SSC112} and by definition it has the same form of
$\langle{\bf S}_i\cdot{\bf S}_j\rangle$. After a Fourier transform, the Hamiltonian is given by:
\begin{eqnarray}
H&=&\sum_{\bf k} \left\{2g\xi\tilde{S}^2(1-\gamma_{\bf k})\phi_{\bf k}\phi_{\bf -k}+\right.\nonumber\\
&&+\left.[2g(1+\gamma_{\bf k})+D]S_{\bf k}^z S_{\bf -k}^z\right\},
\end{eqnarray}
with the structure factor $\gamma_{\bf k}=\frac{1}{2}(\cos k_x+\cos k_y)$.
The Hamiltonian is diagonalized by introducing the canonical transformation:
\begin{subequations}
\begin{eqnarray}
\phi_{\bf k}&=&\frac{1}{\sqrt{2}}\left[\frac{2g(1+\gamma_{\bf k})+D}{2g\xi\tilde{S}^2(1-\gamma_{\bf k})}\right]^{1/4}(a_{\bf k}^\dagger+a_{-{\bf k}})\\
S_{\bf k}^z&=&\frac{i}{\sqrt{2}}\left[\frac{2g\xi\tilde{S}^2(1-\gamma_{\bf k})}{2g(1+\gamma_{\bf k})+D}\right]^{1/4}(a_{\bf k}^\dagger-a_{-{\bf k}}),
\end{eqnarray}
\end{subequations}
where $a_{\bf k}^\dagger$ and $a_{\bf k}$ are boson creation and annihilation operators,
respectively. After a straightforward calculation, we find in the continuous limit:
\begin{equation}
\left\langle\left(\frac{S_i^z}{\tilde{S}}\right)^2\right\rangle_0=\frac{1}{2}\int\frac{\ud^2{\bf k}}{4\pi^2}\sqrt{\frac{2g\xi\tilde{S}^2(1-\gamma_{\bf k})}{2g(1+\gamma_{\bf k})+D}}\coth\left(\frac{\beta E_{\bf k}}{2}\right)
\end{equation}
and
\begin{equation}
\langle\phi_{\bf k}\phi_{-{\bf k}}\rangle_0=\frac{1}{2}\sqrt{\frac{2g(1+\gamma_{\bf k})+D}{2g\xi\tilde{S}^2(1-\gamma_{\bf k})}}\coth\left(\frac{\beta E_{\bf k}}{2}\right),
\end{equation}
in which $E_{\bf k}=2\sqrt{2g\xi\tilde{S}^2(1-\gamma_{\bf k})[2g(1+\gamma_{\bf k})+D]}$
are the eigenvalues of the energy operator, $\langle\ldots\rangle_0$ means a thermal average
calculated through the quadratic Hamiltonian and the integral is evaluated over the first
Brillouin zone. The $\xi$ parameter is given by:
\begin{equation}
\xi=\left\langle\sqrt{1-\left(\frac{S_i^z}{\tilde{S}}\right)^2}\sqrt{1-\left(\frac{S_j^z}{\tilde{S}}\right)^2}\cos(\phi_i-\phi_j)\right\rangle,
\end{equation}
where the exact average is taken by considering the original Hamiltonian (\ref{eq:original_hamiltonian}).
To evaluate the above expression, we have approximated the average by applying the diagonalized harmonic Hamiltonian.
Once $\phi_i$ and $S_i^z$ are uncoupled operators and $\phi_i$
has a Gaussian distribution, we have:
\begin{equation}
\label{eq:correlation}
\xi\cong\left[1-\left\langle\left(\frac{S_i^z}{\tilde{S}}\right)^2\right\rangle_0\right] e^{-\frac{1}{2}\langle(\phi_i-\phi_j)^2\rangle_0}
\end{equation}
with
\begin{equation}
\langle(\phi_i-\phi_j)^2\rangle_0=\int\frac{\ud^2{\bf k}}{2\pi^2}(1-\gamma_{\bf k})\langle\phi_{\bf k}\phi_{-{\bf k}}\rangle_0.
\end{equation}
Equation (\ref{eq:correlation}) is solved self-consistently and the solutions are used to
determine the critical point $D_c$. At zero temperature, a spin-spin correlation between nearest neighboring
spins is finite for $D<D_c$ (the ordered phase) and abruptly vanishes when $D$ tends to
$D_c$, characterizing the quantum phase transition. In the large-$D$ region, the
correlation $\xi$ is null and the system falls into a disordered regime. The critical points are
numerically evaluated and the results are shown in Fig. (\ref{fig:dc_scha}). As found in the
bond operator method in previous section, the behavior of the critical point is almost linear as a function of the biquadratic
constant but its value is around 10$\%$ bigger (compared with the numeric results).

At finite temperatures, the system is disordered but the spin-spin correlation order-parameter has a different
behavior below and above a critical temperature $T_c$. For $0<T<T_c$, the correlation has an algebraic
decay while it falls exponentially for $T>T_c$. It is important to highlight that it is not a transition associated
with a broken symmetry as that considered by Mermin-Wagner theory and so, it is more similar to the
BKT transition. At low temperatures and in the classical
limit, we have:
\begin{equation}
1-\left\langle\left(\frac{S_i^z}{\tilde{S}}\right)^2\right\rangle_0\cong 1-t I
\end{equation}
where the reduced temperature $t=\frac{T}{4g\tilde{S}^2}$ and
\begin{equation}
I=\int\frac{\ud^2{\bf k}}{4\pi^2}\frac{2g\tilde{S}^2}{2g(1+\gamma_{\bf k})+D}
\end{equation}
represents the out-of-plane fluctuations. For the in-plane component we have $\langle(\phi_i-\phi_j)^2\rangle_0\cong \frac{2t}{\xi}$
and the equation (\ref{eq:correlation}) is written as:
\begin{equation}
\label{eq:xiapp}
\xi\cong\left(1-tI\right) e^{-\frac{t}{\xi}}
\end{equation}
The transition temperature is evaluated perfoming a self-consistent calculation of (\ref{eq:xiapp}) and the point
where $\xi$ abruptly goes to zero is taken as $T_c$. However the
calculated values are overestimated. As pointed by Ariosa and Beck\cite{HPA65,PRB54}, a self-consistent harmonic approximation
attributes an excessive energetic cost to topological excitations which reflects the larger transition temperature.
The problem is consider an unique bump centered at $\phi_0=\phi_i-\phi_j=0$ while should be considered a bump 
at each $\phi_n=2\pi n$ due the periodic potential of the lattice. Following Ariosa and Beck, we have added a
correction term for the in-plane component:
\begin{equation}
\label{eq:improved}
\langle(\phi_i-\phi_j)^2\rangle_0\cong \frac{2t}{\xi}+4\pi^2p,
\end{equation}
in which 
\begin{equation}
p=1-\textrm{erf}\left(\sqrt{\frac{\xi g\tilde{S}^2\pi^2}{2T}}\right)
\end{equation}
implements the probability for the phase difference to be out of the interval $[-\pi,\pi]$ ($\textrm{erf}(x)$ is the error function).
\begin{figure}[h]
\centering
\epsfig{file=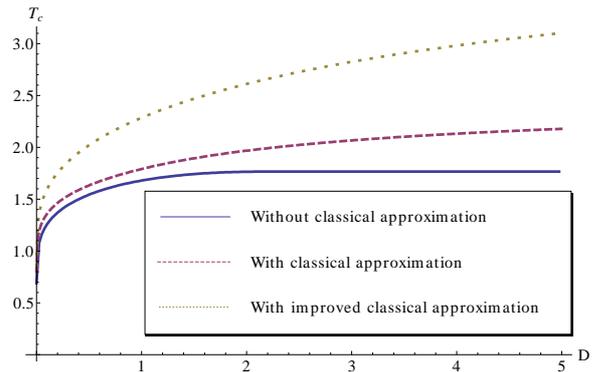,height=60mm}
\caption{The difference between the three methods applied to evaluated the transition temperature for $J_2=0.5$.}
\label{fig:tc_1}
\end{figure}

In Fig. (\ref{fig:tc_1}) we show three different results for the transition temperature (for $J_2=0.5$, in units of $J_1$) 
as function of $D$: obtained by the equation (\ref{eq:xiapp}), 
using the improved form (\ref{eq:improved}) and numerically evaluated through the equation (\ref{eq:correlation}) 
(without the classical limit assumption). How we can see, the classical approximation overestimates the transition temperature
but the correction implemented by Ariosa and Bech provides a better according with the equation (\ref{eq:correlation}).
\begin{figure}[h]
\centering
\epsfig{file=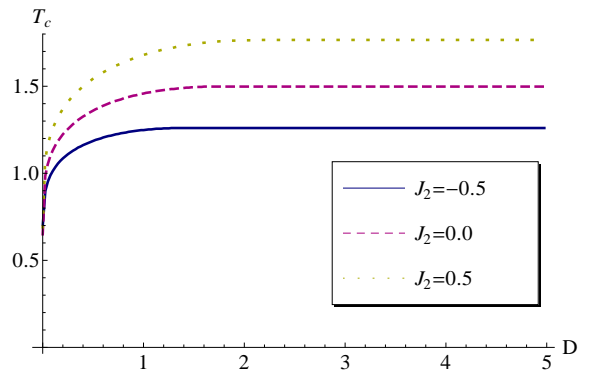,height=60mm}
\caption{The transition temperature as a function of $D$ (in units of $J_1$).}
\label{fig:tc_d}
\end{figure}

Already in Fig. (\ref{fig:tc_d}) we plot the results obtained for $T_c$ as a function of the biquadratic constant for some values of $J_2$ using only the third method. 
We can note that the transition temperature assumes a constant value for $D\gtrsim1.5$ for all values of the biquadratic constant. The critical
temperature has the approximated limit $0.6$ when $D$ tends to zero, independently of the constant $J_2$.
On the other hand, close to large-$D$ phase, the critical temperature abruptly goes to zero when $D\to D_c^-$ once the system is
disordered even at zero temperature in this phase.

\begin{figure}[h]
\centering
\epsfig{file=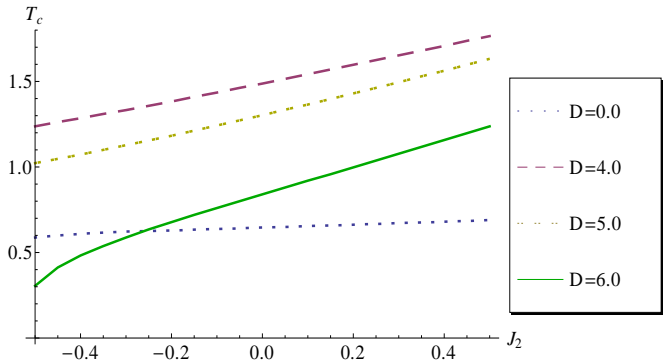,height=50mm,width=90mm}
\caption{The transition temperature as a function of $J_2$ (in units of $J_1$).}
\label{fig:tc_j2}
\end{figure}

Figure (\ref{fig:tc_j2}) shows the relation between the critical temperature and the biquadratic
constant. For $D<D_c$, $T_c$ presents an almost linear increasing with $J_2$ and for $D=0$ we can
see an almost constant value as shown Fig.{\ref{fig:tc_d}}.

\section{Conclusion}
\label{conclusion}
We have studied the Anisotropic Biquadratic Heisenberg Model using different techniques.
In special we have analysed the quantum phase transition associated with the single-ion
anisotropy at zero temperutre and a thermal phase transition in the disordered regime ($T>0$). 
At zero temperature and for small values of $D$ ($D<D_c$),
the system is in a gapless ordered state while for $D>D_c$ there is no long-range spin order ($m=0$). Using the
$SU(2)$ Schwinger formalism, we have found a decreasing behavior for the condensate density,
which indicates a phase transition to large $D$ although it can not be exactly determined by this method. We
have shown also the existence of a nematic phase even in large-D phase. Analyzing the large-$D$ region (where
the ground state is gapped and without magnetization, $m=$) we have determined the critical points $D_c$ as a function
of the biquadratic constant $J_2$. The critical values of $D_c$ have an almost linear behavior with $J_2$ and
within the interval $-1\leq J_2\leq 1$, there is always a quantum phase transition.
Finally, applying the SCHA method, we have obtained similar results
for the critical points and also, the transition temperature between a disordered phase with algebraic decay
and the regime with exponential decay of the order-parameter correlation, including the dependence with $D$ and
$J_2$.

\section*{Acknowledgment}
This work was supported by Funda\c c\~ao de Amparo \`a Pesquisa do estado de Minas Gerais
(FAPEMIG) and Conselho Nacional de Desenvolvimento Científico e Tecnológico (CNPq), Brazil.

\bibliography{abhm.bib}

\end{document}